\documentclass[aps,prb,superscriptaddress,preprint]{revtex4}

\usepackage{graphicx}
\usepackage{subfig}
\usepackage{url}

\begin{document}

\title{\ACEMD: Accelerating bio-molecular dynamics in the microsecond time-scale}

\author{M. J. Harvey}
\email{m.j.harvey@imperial.ac.uk}
\affiliation{Information and Communications Technologies,\\ Imperial College London, South Kensington, London, SW7 2AZ, UK}
\author{G. Giupponi}
\email{giupponi@ffn.ub.es}
\affiliation{Department de Fisica Fundamental, Universitat de Barcelona, Carrer Marti i Franques 1, 08028 Barcelona, Spain}
\author{G. De Fabritiis}
\email{gianni.defabritiis@upf.edu}
\affiliation{Computational Biochemistry and Biophysics Lab (GRIB-IMIM), Universitat Pompeu Fabra, Barcelona Biomedical Research Park (PRBB), C/ Doctor Aiguader 88, 08003 Barcelona, Spain}

\newcommand{\NVIDIA}{Nvidia\ }
\newcommand{\MSHAKE}{M-shake\ }
\newcommand{\ACEMD}{ACEMD\ }
\newcommand{\CELLMD}{CellMD\ }

\begin{abstract}
The high arithmetic performance and intrinsic parallelism of recent graphical
processing units (GPUs) can offer a technological edge for molecular dynamics
simulations.  \ACEMD is a production-class bio-molecular dynamics (MD)
simulation program designed specifically for GPUs which is able to achieve
supercomputing scale performance of $40$ nanoseconds/day for all-atom protein
systems with over $23,000$ atoms. We illustrate the characteristics of the code,
its validation and performance.  We also run a microsecond-long trajectory for 
an all-atom molecular system in explicit TIP3P water on a single workstation
computer equipped with just $3$ GPUs. This performance on cost effective hardware allows 
\ACEMD  to reach microsecond timescales routinely with important implications in terms of scientific applications. 
\end{abstract}
\maketitle

\section{Introduction}

The simulation of mesoscopic scales (microseconds to milliseconds) of
macromolecules continues to pose a challenge to modern computational
biophysics. Whilst the fundamental thermodynamic framework behind the
simulation of macromolecules is well characterised, exploration of biological time
scales remains beyond the computational capacity routinely available to many researchers.
This has significantly inhibited the widespread use of molecular
simulations for \textit{in silico} modelling and prediction \cite{ddt}.

Recently, there has been a renewed interest in the development of molecular
dynamics simulation techniques. D. E. Shaw Research \cite{anton} has fostered
several significant algorithmic improvements including mid-point \cite{midpoint} and neutral-territory methods
 \cite{nt} for the summation of non-bonded force calculations, a new
molecular dynamics package called Desmond \cite{Desmond} and Anton
\cite{anton}, a parallel machine for molecular dynamics simulations that uses
specially-designed hardware. 
Other parallel MD codes, such as Blue matter \cite{fitch06}, NAMD \cite{namd} and Gromacs4 \cite{GROMACS4},
have been designed to perform parallel MD simulations
across multiple independent processors, but latency and bandwidth limitations in the
interconnection network between processors reduces parallel scaling unless the
size of the simulated system is increased with processor count. Furthermore, dedicated, highly-parallel machines are 
usually expensive and not reservable for long periods of time due to cost constraints  and allocation restrictions. 

A further line of development of MD codes consists of using commodity high
performance accelerated processors \cite{ddt}. This approach has become an
active area of investigation, particularly in relation to the Sony-Toshiba-IBM
Cell processor\cite{gdfcell} and graphical processing units (GPUs).  Recently, De Fabritiis
\cite{gdfcell} implemented an all-atom biomolecular simulation code, CellMD,
targeted to the architecture of the Cell processor (contained
within the Sony Playstation3) that reached a sustained performance of $30$
Gflops with a speedup of $19$ times compared to the single CPU version of the
code. At the same time, a port of the Gromacs code for implicit solvent models
\cite{luttmann2008amd} was developed and used by the Folding@home distributed
computing project\cite{FAH} on a distributed network of Playstation3s.
Similarly, \CELLMD was used in the PS3GRID.net project \cite{ps3grid} based on
the BOINC platform \cite{BOINCweb} moving all-atom MD applications into a
distributed computing infrastructure.

Pioneers in the use of GPUs for production molecular dynamics
\cite{FAH} had several limitations imposed by the restrictive,
graphics-orientated OpenGL programming model\cite{opengl} then available.  In
recent years, commodity GPUs have acquired non-graphical,
general-purpose programmability and have undergone a doubling of computational
power every $12$ months, compared to $18-24$ months for traditional CPUs
\cite{ddt}. Of the devices currently available on the market, those produced by
\NVIDIA offer the most mature programming environment, the so-called compute
unified device architecture (CUDA)\cite{cuda}, and have been the focus of the
majority of investigation in the computational science field. 

Several groups have lately shown results for MD codes which utilise CUDA-capable 
GPUs. Stone \textit{et al} \cite{cionize} demonstrated the
GPU-accelerated computation of the electrostatic and van der Waals forces,
reporting a $5.4$ times speed-up with respect to a conventional CPU. Meel
\textit{et al} \cite{meel08} described an implementation for simpler
Lennard-Jones atoms which achieved a net speedup of up to $40$ times over a
conventional CPU. Unlike the former, the whole simulation is performed on the
GPU. Recently, Phillips \textit{et al} have reported experimental GPU-acceleration of NAMD \cite{namdgpu} yielding speedups of up to $7$ times over NAMD 2.6.
 
In this work, we report on a molecular dynamics program called \ACEMD  which is
optimised to run on  \NVIDIA  GPUs and which has been developed with
the aim  of advancing the frontier of molecular simulation towards the ability
to routinely perform microsecond-scale simulations. \ACEMD  maximises
performance by running the whole computation on the GPU rather than offloading
only selected computationally-expensive parts. We have developed \ACEMD  to
implement all features of a typical MD simulation
 including those usually required for production simulations such as
particle-mesh Ewald (PME\cite{pme}) calculation of long range electrostatics,
thermostatic control  and bond constraints. 
The default force-field format used by \ACEMD  is CHARMM \cite{charmm27} although
 the code can also use the Amber99 \cite{forceamber} force field once converted to
the format of the former\cite{portamber}. \ACEMD  also provides a scripting interface to
control and program the molecular dynamics run to perform complex protocols
like umbrella sampling, steered molecular dynamics and sheared boundary
conditions.  


\section{ GPU architecture} 

\begin{figure}
\begin{center}
\includegraphics[width=7.5cm]{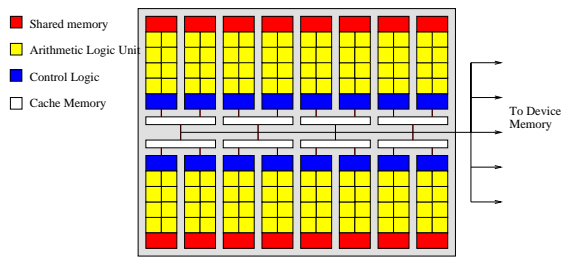}
\caption{\NVIDIA GPU design is based around an 8 core single program, multiple
data (SPMD) processor. Each core has local storage provided by the register
file and access to a shared memory region. Read/write access to the main device
memory is uncached, except for some specific read-only access modes.
The above figure represents the G80-series device with 16 such processors, whilst the contemporary G200 contains 30 (Tesla C1060,
GTX280), giving 240 cores.} 
\label{fig:filler}
\end{center}
\end{figure}

The G80 and subsequent G200 generations of \NVIDIA  GPU architectures are
designed for data-parallel computation, in which the same program code is
executed  in parallel on many data elements. The CUDA programming model, an
extended C-like language for GPUs, abstracts the implementation details of the
GPU so that the programmer may easily write code that is portable between
current and future GPUs. \NVIDIA  GPU devices are implemented as a set of
multiprocessor (MP) devices, each of which is capable of synchronously executing $32$ program
threads in parallel (called warp) and managing up to $1024$ concurrently (Figure
\ref{fig:filler}).  Current \NVIDIA  products based on these devices are able
to achieve up to $933$ Gflops in single precision.  By comparison, a contemporary quad core Intel
Xeon CPU is capable of approximately 54 Gflops \cite{dongarra}. A brief
comparison of the characteristics of these devices is given in Table
\ref{table:comparison}.  Each MP has a set of $32$ bit registers, which are
allocated as required to individual threads, and a region of low-latency shared
memory that is accessible to all threads running on it. The MP is able to
perform random read/write access to external memory. Access to this global
memory is uncached and so incurs the full cost of the memory latency (up to $400$ cycles). However, when accessed via the GPU's texturing units,
reads from arrays in global memory are cached, mitigating the impact of global
memory access for certain read patterns. Furthermore, the texture units are
capable of performing linear interpolation of values into multidimensional (up to
3D) arrays of floating point data.

Whilst the older G80 architecture supported only single-precision IEEE-754
floating point arithmetic, the newer G200 design also supports double-precision
arithmetic, albeit at a much lower relative speed. The MP has special hardware
support for reciprocal square root, exponentiation and trigonometric functions,
allowing these to be computed with  low latency but at the expense of slightly
reduced precision. 

Program fragments  written to be executed on the GPU are known as
\textit{kernels} and are executed in \textit{blocks}.  Each block consists of multiple
instances of the kernel, called \textit{threads}, which are run concurrently on
a single multiprocessor. The number of threads in a block is limited by the
resources available on the MP but multiple blocks may be grouped together
as a \textit{grid}.  The CUDA runtime, in conjunction with the GPU hardware
itself, is responsible for efficiently scheduling the execution of a grid of
blocks on available GPU hardware. CUDA does not presently provide a mechanism
for transparently using multiple GPU devices for parallel computation.  For
full details of the CUDA environment, the reader is referred to the SDK documentation \cite{cudasdk}. 

\begin{table}
\begin{tabular}{|l|lll|}
\hline
  & Tesla & Tesla& Intel \\
		& C870 (G80)  & C1060 (G200)       & Xeon 5492 \\
\hline
Cores       & 128 & 240 & 4 \\
Clock (GHz) & 1.350 & 1.296 & 3.4 \\
\small{Mem bandwidth (MB/s)}     &  77 & 102 & 21 \\
Gflops      & 512 (sp) & 933 (sp) 78 (dp) & 54 (dp) \\
Power (W)      & 171 & 200 & 150 \\
Year        & 2007 & 2008 & 2008 \\
\hline
\end{tabular}
\caption{Summary of characteristics of first and second generation \NVIDIA GPU
compute devices, with a contemporary Intel Xeon shown for comparison. Data taken
from manufacturers' data sheets. (sp stands for single precision and dp for double precision).}

\label{table:comparison}
\end{table}



\section{Molecular dynamics on the GPU} 

\ACEMD  implements all features of an MD simulation on a
CUDA-compatible GPU device, including those usually required for production
simulations in the NVT ensemble (\textit{ie} bonded and non-bonded force term
computation, velocity-Verlet integration, Langevin thermostatic control, smooth Ewald
long range electrostatics (PME)\cite{pme,spme} and hydrogen bond constraints).
Also implemented is the hydrogen mass repartitioning scheme described in
Ref.  \cite{feenstra99} and used for instance in codes such as Gromacs, which allows an increased timestep of up to $4$ fs.
The code does not presently contain a barostat, so simulations in the NPT
ensemble are not possible. However, it is noted that with large molecular systems,
changes in volume due to the pressure control are very limited after an initial
equilibration making NVT simulations viable
for production runs. \ACEMD  supports the CHARMM27 force field and
Amber99 in CHARMM format \cite{portamber}, PDB, PSF and DCD file formats \cite{MDX} as well as
steered molecular dynamics \cite{gramicidin}, check-pointing and input files
compatible with a widely used MD codes such as NAMD.

The computation of the non-bonded force terms dominates the computational cost
of MD simulations and it is therefore important to use an efficient algorithm.
As in \cite{gdfcell,meel08}, we implement a cell-list scheme in which particles
are binned according to their co-ordinates.  On all-atom biomolecular systems a cutoff of
$R=12$ {\AA} with bins $R/2$ gives an average cell population of approximately
$22$ atoms. This is comparable to the warp size of $32$ for current \NVIDIA
GPUs. In practice, however, transient density fluctuations can lead to the cell
population exceeding the warp size.  Consequently, the default behaviour of
\ACEMD  is to assume a maximum cell population of $64$.
  The code may also accommodate
a bin size of $R$ for coarse-grained simulations.  The cell-list construction
kernel processes one particle per thread, with each thread computing the cell
in which its atom resides.  To permit concurrent manipulation of a cell-list
array, atomic memory operations are used.

\begin{figure}
\begin{center}
\includegraphics[width=7cm,angle=-90]{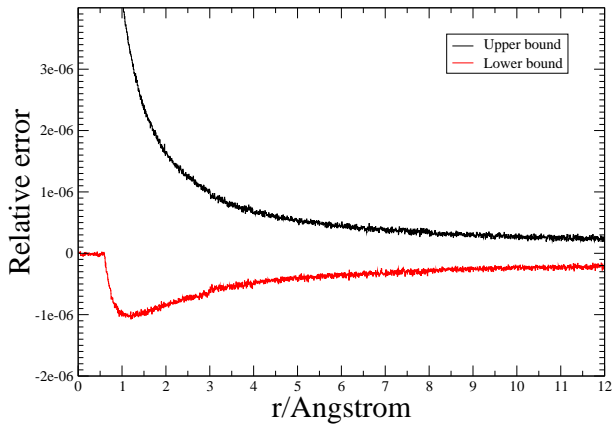}
\caption{The bounds (running average) of the relative error 
$\frac{E_{interp}-E_{calc}}{E_{calc}}$ between directly calculated and  lookup
table (linear interpolation, $n=4096$) values of the van der Waals potential.
Error has a period of $\frac{R_{max}}{n}$} 
\label{fig:potentialdelta}
\end{center}
\end{figure}

The non-bonded force computation kernel processes a single cell per thread
block, computing the full Lennard-Jones and electrostatic force on each
particle residing within it. All of the cells within $R$ of the current cell
(including the a copy of the cell itself) are loaded into shared memory in
turn. Each thread then computes the force on its particle by iterating over the
array in shared memory. In contrast to CPU implementations, reciprocal forces
are not stored for future use (\textit{ie} the force term $F_{ij}$ is not saved for
reuse as $F_{ji}$), because of the relatively high cost of global memory
access.  
The texture units are used to assist the calculation of the electrostatic and
van der Waals terms by providing linearly interpolated values for the radial
components of those functions from lookup tables. The interpolation error is
low and does not affect the energy conservation properties of NVE simulations.
In production runs (Figure \ref{fig:potentialdelta}), the relative force error compared to a reference
simulation performed in double precision is consistently less than $10^{-4}$,
below the $10^{-3}$ error considered the maximum acceptable for biomolecular
simulations\cite{Desmond}.
Particle-mesh Ewald (PME) \cite{spme} evaluation of long-range electrostatics is also
supported by a dedicated kernel. All parts of this computation are performed on the GPU, with support
from the \NVIDIA  FFT library\cite{cufft}. For PME calculations, a cutoff of
$R=9.0$ {\AA}    is considered to provide sufficient accuracy,
 permitting the maximum cell population to be limited to 32 atoms.

To support CHARMM and AMBER force fields, it is necessary to selectively
exclude or scale non-bonded force terms between atoms that share an explicit
bond term. The indices of  excluded and 1-4 scaled pairs are stored in bitmaps,
allowing any pair of particles with indices $i,j$ such that $|i-j|\le 64$ to be
excluded or scaled \cite{cionize}. Exclusions with larger index separations are
also supported in order to accommodate, for example, disulphide bonds but the
additional book-keeping imposes a minor reduction in performance.  Because the
atoms participating in bonded terms are spatially localised, it is necessary
only to make exclusion tests for interactions between adjacent cells despite,
for cells of $R/2$, the interaction halo being two cells thick. A consequent
optimisation is the splitting of the non-bonded force kernel into two versions,
termed \textit{inner} and \textit{outer}, which respectively include and omit
the test. 

Holonomic bond constraints are implemented using the \MSHAKE
algorithm\cite{mshake} and RATTLE for velocity constraints \cite{rattle} within
the velocity Verlet integration scheme \cite{lfvv}.  \MSHAKE  is an iterative
algorithm and, in order to achieve acceptable convergence it is necessary to
use double precision arithmetic (a capability available only on
G200/architecture 1.3 class devices).  For the pseudo-random number source for
the Langevin thermostat we use a Mersenne twister kernel, modified from the
example provided in the CUDA SDK.




\section{Single-precision floating-point arithmetic validation}

\ACEMD  uses single-floating point arithmetic because the performance of GPUs
on single precision is much higher than double precision and the limitation of
single floating-point can be controlled well for molecular
dynamics\cite{GROMACS4,gdfcell}.  Nevertheless, we validate in this section
the conservation properties of energy in a NVT simulation using rigid and
harmonic bonds, as constraints have shown to be more sensitive to numerical
precision. 
Potential energies were checked against NAMD values for the initial configuration of a
set of systems, including disulphide bonds, ionic system, protein and membranes,
in order to verify the correctness of the force calculations by assuring that
energies were identical within $6$ digits. The Langevin thermostat algorithm
was tested for three different damping frequencies $\gamma = 0.1, 0.5, 1.0$ with
a reference temperature of $T = 300K$ and both with and without constraints.

\begin{table}
\begin{tabular}{|c|c|c|c|}
\hline
Timestep (fs) &  constraints  & HMR &  $K_bT$/ns/dof \\
\hline
1 & no & no &   0.00021    \\
2 & yes & no &  -0.00082   \\
4 & yes & yes & -0.00026   \\
\hline
\end{tabular}
\caption{
\label{table:conservation}
Energy change in the NVE ensemble per nanosecond per degrees of freedom (dof) in $K_bT$ units
for  dihydrofolate reductase (DHFR) using different
integration timesteps, constraints and hydrogen mass repartitioning (HMR) schemes.
}
\end{table}
The test simulations consist of  nanosecond runs of dihydrofolate
reductase (DHFR) joint AMBER-CHARMM benchmark with volume $62.233\times62.233\times62.233$ ${\AA}^3$  (a total of 23,558
atoms) \cite{Desmond}.  Each simulation system was firstly equilibrated at
a temperature of $T=300K$ and then relaxed in the NVE ensemble. A 
reference simulation  with harmonic bonds and timestep $dt=1$ fs was also performed, as well as
simulations with $dt=2$ fs  using
rigid constraints and with $dt = 4$ fs with rigid constraints and hydrogen mass
repartitioning (HMR) with a factor $4.0$, as in
\cite{HMASSREF}.  In Table \ref{table:conservation}, we show the energy change per
nanosecond per degrees of freedom in units of $K_bT$, which is similar with other single and double precision
codes MD\cite{lippert07}. 
We note that even when using bigger timesteps and a
combination of \MSHAKE and hydrogen mass repartitioning, energy conservation is
reasonably good, and much slower than the timescale at which the thermostat would act. 
Hydrogen mass repartitioning is an elegant way to increase the timestep up to
$4$ fs by increasing the momentum of inertia of groups of atoms bonded to
hydrogen atoms.  The mass of the bonded heavy atoms to hydrogens is
repartitioned among hydrogen atoms, leaving  the total mass of the
system unchanged.  As individual atom masses do not appear in the expression for the
equilibrium distribution, this repartition affects only the dynamic properties
of the system not the equilibrium distribution. Following \cite{feenstra99}, a
factor $4$ for hydrogens affects only marginally the diffusion and viscosity of
TIP3P water (which is in any case inaccurate when compared to experimental data). A similar speed
up could also be obtained by using a smaller timestep with the evaluation of the
long range electrostatic terms every other timestep. 
 


\begin{figure}
\begin{center}
\includegraphics[width=7cm]{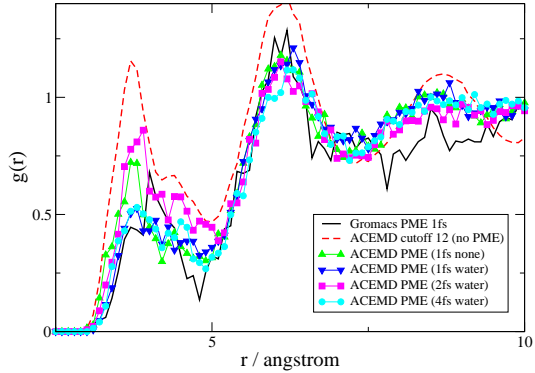}
\caption{Plot of Na-Na pair distribution functions for a 1M NaCl water box
  as in \cite{grf}.}
\label{fig:PME}
\end{center}
\end{figure}

We also validated the implementation of the PME algorithm to compute long range
electrostatics forces. We run a set of simulations using different timesteps
and algorithms as above ($dt=1,2$ fs rigid bonds , $dt=4$ fs rigid bonds and hydrogen
mass repartitioning) on a $40.5\times40.5\times40.5$\AA$^3$ box of $1$ M solution of NaCl in water
(6461 atoms), as in \cite{grf}. PME calculations were performed with a $64\times64\times64$
grid size.  Two simulations of the same system were used as reference, one  with Gromacs
\cite{GROMACS4} with PME as in \cite{grf}, and the other using \ACEMD with an
electrostatic cutoff of $12$ {\AA} without PME.
We calculated the Na-Na pair distribution function $g(r)$ in Figure \ref{fig:PME}
in order to compare the simulation results for different simulations and
methods, as from \cite{grf} Na-Na $g(r)$ results as the quantity more sensitive to different
methods for electrostatics calculations.  We note that for all integration
timesteps used, \ACEMD  agrees well with the reference simulation made
with Gromacs. In addition, using PME gives consistently better results than
using a $12$ {\AA} cutoff for this simple homogeneous system, as expected. 
A direct validation of the pair distribution function with the hydrogen mass
repartitioning  method is also shown in Fig. \ref{fig:PME} comparing the $g(r)$ for 
timestep equal $1,2,4$.

\section{Performance}

\begin{table}
\begin{tabular}{|l|l|l|}
\hline
Program & CPU Cores \& GPUs & ms/step \\
\hline
\ACEMD  & 1 CPU, 1 GPU (240 cores)  & 17.55 \\
\ACEMD  & 3 CPU, 3 GPU (720 cores)  & 7.56  \\
\hline
NAMD2.6    & 128 CPU (64 nodes)  & 9.7  \\
NAMD2.6    & 256 CPU (128 nodes)  & 7.0  \\
\hline
Desmond & 32 CPU (16 nodes)   & 11.5   \\
Desmond & 64 CPU (32 nodes)   & 6.3   \\
\hline
Gromacs4 & 20 CPU (5 nodes)   & 7    \\
\hline

\end{tabular}
\caption{
\label{table:scaling1}
Performance of \ACEMD  on the DHFR benchmark. GPUs are
\NVIDIA  GTX 280. NAMD, Desmond and Gromacs performances are indicative of the
orders of magnitude speed-up obtained with GPUs and  \ACEMD  as they are all
performed on different CPU systems (from Refs \cite{Desmond,GROMACS4}, Gromacs figures interpolated from Figure 6 of \cite{GROMACS4}).
}
\end{table}

\begin{table}
\begin{tabular}{|l|l|l|}
\hline
Program & CPU Cores \& GPUs & ms/step \\
\hline
\ACEMD & 1 CPU, 1 GPU  (1 node) & 73.4 \\
\ACEMD & 3 CPU, 3 GPU  (1 node) & 32.5 \\
NAMD   & 4 CPU, 4 GPU (1 node)  & 87   \\
NAMD   & 16 CPU, 16 GPU (4 nodes)& 27  \\
NAMD   & 60 CPU  (15 nodes)     & 44   \\
\hline
\end{tabular}
\caption{
\label{table:scaling2}
Performance of \ACEMD and NAMD on the apoA1 benchmark. \ACEMD run using \NVIDIA
GTX 280 GPUs ($R=9$\AA, PME every step), NAMD ($R=12$\AA, PME every 4 steps)
run with G80-series GPUs (approximately half as fast)  NAMD performance data
taken from \cite{namdgpu}. 
}
\end{table}

The current implementation of \ACEMD  is parallelized in a 
task parallel manner designed to scale across just $3$ GPUs attached to a single host
system. A simple force-decomposition scheme\cite{force-decomp-plimpton} is
used, in which each GPU computes a subset of the force terms. These force terms
are summed on by the host processor and the total force matrix transferred back
to each GPU which then perform integration of the whole system. \ACEMD
dynamically load-balances the computation across the GPUs. This allows the
simulation of heterogeneous molecular systems and also accommodates variation
due to host system architecture (for example, different speed GPUs or GPU-host
links). For simulations requiring PME, a heterogeneous task decomposition is
used, with a subset of GPUs dedicated to PME computation.

The performance benchmark is based on the DHFR molecular system with a cutoff
of $R=9$\AA, switched at $7.5$\AA, $dt=4$ fs, PME for long range electrostatic
with $64 \times 64 \times 64$ grid size and fourth order interpolation, \MSHAKE
constraints for hydrogen bonds and hydrogen mass repartitioning. All simulations
were run on a PC equipped with $4$ \NVIDIA  GPU GTX 280 cards at 1.3GHz (just 3
GPUs used for these tests), a quad core AMD Phenom processor (2.6GHz), MSI
board with AMD790 FX chipset, 4GB RAM running Fedora Core 9, CUDA toolkit 2.0
and the \NVIDIA  graphics driver 177.73.  Performance results  reported
in Table \ref{table:scaling1} indicates that \ACEMD  requires $17.55$ ms
per step with the DHFR system and $7.56$ ms per step
when run in parallel over the $3$ GPUs.  As expected by the simple task
decomposition scheme, \ACEMD  achieves a parallel efficiency of 2.3 over $3$ GPUs.
Further device-to-device communication directives
 may substantially improve
these results as they will enable the use of spatial-decomposition  parallelisation
strategies, such as  neutral territory (NT) schemes\cite{nt}.
Comparing directly the maximum performance of \ACEMD  on the DHFR system with
results of various MD programs from Ref. \cite{Desmond} we obtain a performance
approaching that of  $256$ CPU cores using NAMD and $64$ using Desmond on a cluster with
fast interconnect. Using hydrogen mass repartitioning and a time step of
$4$ fs integration timestep it is possible to simulate trajectories of over $45$ ns
per day with $3$ GPUs and almost $20$ nanosecond per day with a single
GPU. An highly optimised code such as Gromacs4 requires only $20$ CPU cores to
deliver similar performance to $3$ GPUs on DHFR \cite{GROMACS4}, but the
calculations are not identical as, for instance, there are several optimisations applied to
water.

Representative performance data for \ACEMD and the the GPU-accelerated version
of NAMD \cite{namdgpu} for the apoA1 benchmark system (92,224 atoms) is given in Table
\ref{table:scaling2}. Differences between the simulation and hardware
configurations prevent a direct comparison but it is salient to note that
because the enhanced NAMD  retains the spatial decomposition parallelism it is
able to scale across multiple GPU-equipped hosts, whilst \ACEMD is designed for
optimal performance on a small number of GPUs.

\section{Microsecond simulations on workstation hardware}

\begin{figure}
\begin{center}
(a)\includegraphics[width=7cm]{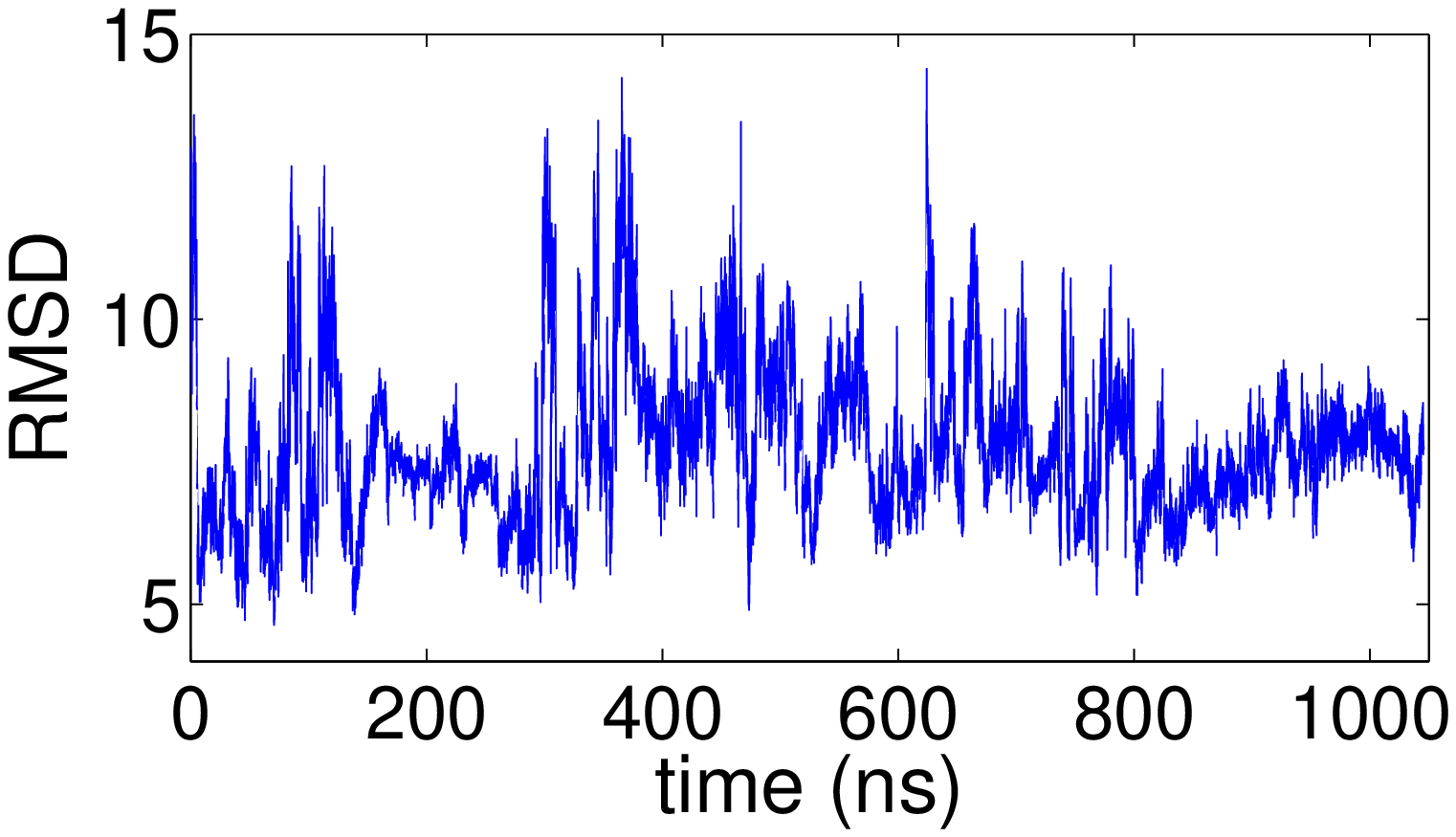}\\
(b)\includegraphics[width=3.5cm]{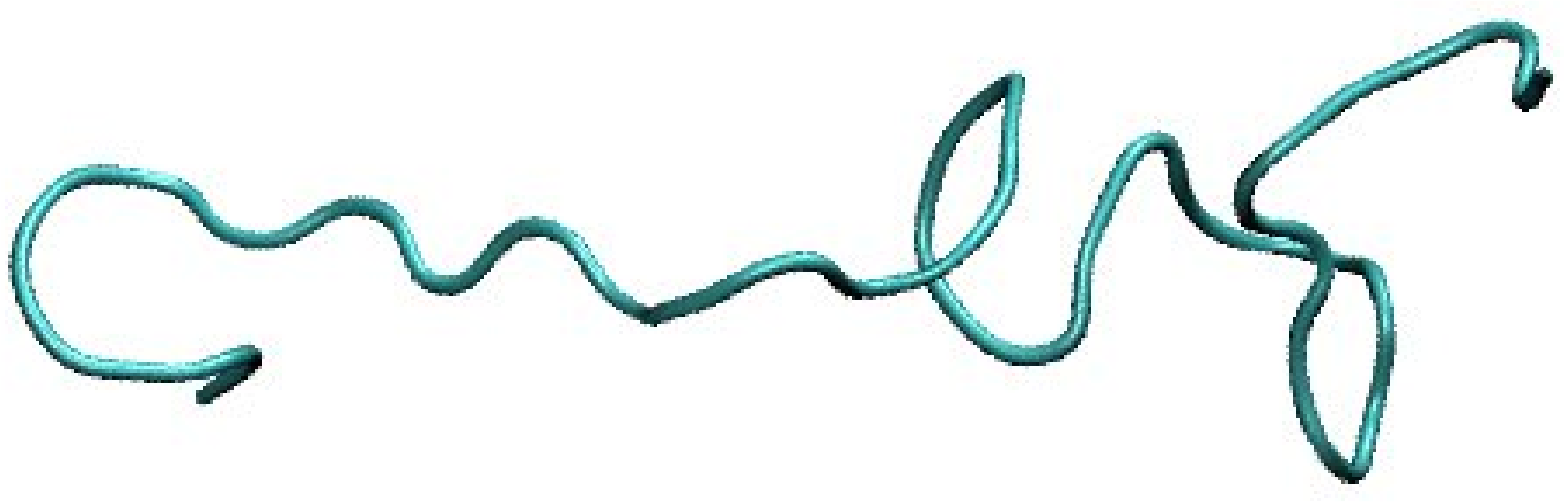}
(c)\includegraphics[width=3cm]{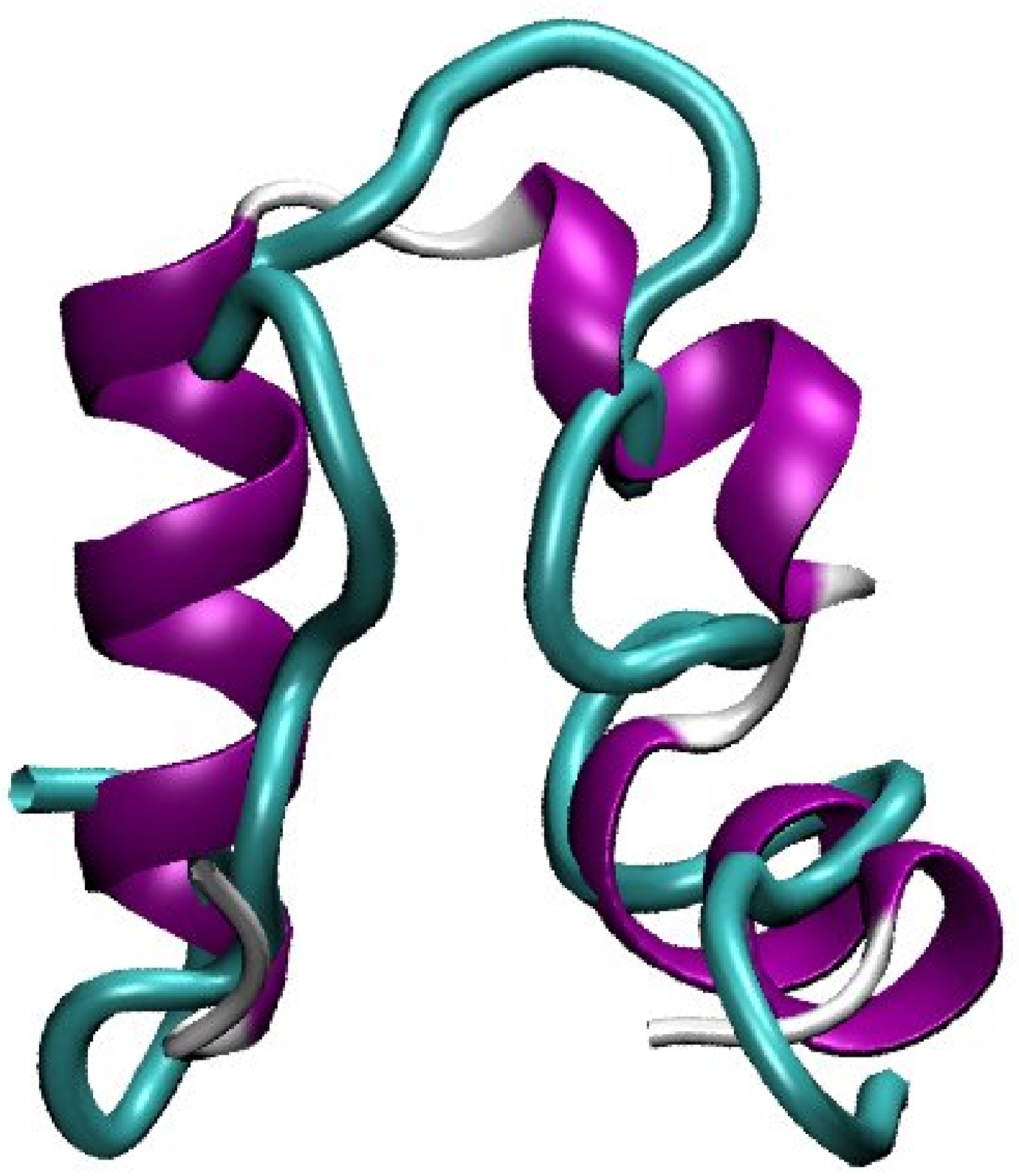}
\end{center}
\caption{(a) The RMSD of the backbone of the protein during the microsecond
simulation starting from the unfolded configuration (b) of the Villin system with 13,701 atoms (TIP3P water not shown for clarity). 
Within our simulation window the minimum RMSD was 4.87 {\AA} for which the resulting best structure is overlapped
with the crystal structure in (c).}
\label{fig:rmsd}
\end{figure}

To provide a direct demonstration that molecular simulations have now entered the
microsecond regime {\it routinely} we perform  a microsecond long trajectory
performed on a workstation-class PC.  
We use for this task the chicken Villin headpiece (HP-35) structure, 
one of the smallest polypeptides with a stable globular structure comprising
three alpha-helices placed in a ''U``-shaped form, as shown in Figure
\ref{fig:rmsd}c. Due to its small size, it is commonly used as a subject in
long molecular simulations for folding studies, for instance\cite{ensign2007}, which
uses highly parallel distributed-computing to compute many trajectories to
fully sample the phase space of the folding process.  Alternatively,
mutanogesis studies on folding \cite{piana2008} have been performed using biased
methods to accelerate the sampling. 

The  Villin headpiece (PDB:1YRF) was fully solvated in TIP3P water and Na-Cl at
150mM (a total of $13701$ atoms) using the program VMD \cite{VMD} and the
CHARMM force-field. The system was then equilibrated at 300K and 1atm for 
$10$ns using NAMD2.6\cite{namd} with a cutoff of $9$ {\AA}, PME
with a $48\times48\times48$ grid, constraints for all H bond terms and a
timestep of $2$fs. Simulations with \ACEMD were  performed using an NVT ensemble,
hydrogen mass repartitioning and timestep of $4$fs.  Starting from the
final equilibrium configuration of NAMD, we run \ACEMD
at 450K for $40$ ns until the system was completely unfolded (movie available
at \footnote{http://www.vimeo.com/2505856}).  The resulting extended
configuration (Figure \ref{fig:rmsd}b) was then used as the starting point of a
microsecond long single trajectory at a temperature of $305$K. Figure
\ref{fig:rmsd}a, shows the RMSD of the backbone of the protein along the
trajectory. The minimum RMSD was $4.87$. The protein seems to sample quite often
the overall shape of the crystal structure yet not converge towards it
(Figure \ref{fig:rmsd}c).  As this structure is expected to fold in 4-5 microseconds, we
plan to extend the dynamics in the future along with any newer and faster
version of \ACEMD (for instance using the new \NVIDIA GTX295 cards or, more likely, 
quad GPU Tesla S1075 units).  An important consideration with regards to the force-field: with
molecular simulations approaching the microseconds, it is clear that the
accuracy of the force-fields will become more and more important. 
In particular, this system has shown to be very sensitive to the force-field used\cite{piana2008} 
(CHARMM seems to converge poorly towards the folded structure).

\begin{figure}
\includegraphics[width=7cm,angle=-90]{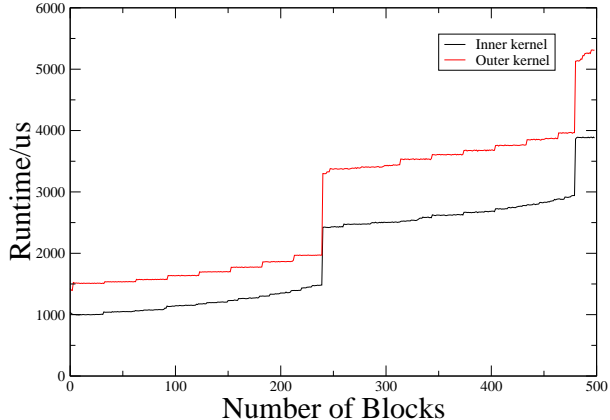}
\caption{Runtime for the non-bonded force calculation kernels on a water
box  as a function of the number of blocks per invocation and run on an \NVIDIA
Tesla C1060 GPU (30 multiprocessors). The inner and outer kernels both have an
occupancy of 8 blocks/multiprocessor. Blocks are distributed across multiprocessors, with the small step increases
indicating an increment in the number of simultaneous blocks. The large steps
indicate the device is fully populated with blocks and that some MPs must
sequentially process further block. Optimal resource usage occurs immediately
before these steps. The effect of gradual divergence between multiprocessors is
seen as block count increases.
  The minimum runtime for the fully parallel case would be 3.4 ms.
} 
\label{fig:scalinglimit}
\end{figure}

\begin{table}
\begin{tabular}{|l|l|l|}
\hline
System & CPUs \& GPUs   & ns/day \\
\hline
DHFR & 1 CPU, 1 GPU  (240 cores)    & 19.7  \\
DHFR & 1 CPU, 3 GPUs  (720 cores)    & 45.7  \\
apoA1 & 1 CPU, 1 GPU  (240 cores)   & 4.6 \\
apoA1 & 3 CPU, 3 GPUs  (720 cores)   & 10.6 \\
Villin  & 3 CPU, 3 GPUs (720 cores)  &  66.0 \\ 
\hline
\end{tabular}
\caption{
\label{table:nanospeed}
Performance of \ACEMD  on the DHFR, apoA1 benchmark and Villin test on $1$ and $3$ GPUs up to 720 cores. 
\ACEMD run using \NVIDIA GTX 280 GPUs on a real production runs ($R=9$\AA, PME every step, time step $4$ fs, constraints and langevin thermostat).}
\end{table}

The production run on a PC equipped with \ACEMD and $3$ \NVIDIA GPUs ($720$ cores) 
required approximately $15$ days ($66$ nanoseconds/day) (see Table \ref{table:nanospeed}) 
and probably represents the limit for current hardware and software implementation,
while $5$ microseconds should be obtainable in the near future using a 4-way
GTX295 based system with $8$ GPU cores. Using
currently-available commodity technology, the construction of computer systems
with up to $8$ directly-attached GPUs has been demonstrated
\footnote{\textit{FASTRA GPU SuperPC}, http://fastra.ua.ac.be/en/index.html,
University of Antwerp (accessed 17th Nov 2008)} \cite{cionize}. GPUs attach to
the host system using the industry-standard PCI-Express interface \cite{pcie}.
This interface is characterised by a bandwidth comparable to that of main
system memory (up to 8GB/s for 16 lane PCIe 2.0 links typically used by
graphics cards) but with a relatively higher latency. 

The GPU resource requirements of the non-bonded kernel make it possible for up
to $8$ independent blocks to be processed simultaneously per multiprocessor.
The limit of parallelization for the execution of the non-bonded kernel occurs
when all blocks may be processed simultaneously by the available
multiprocessors.  Thus, for instance, a cubic simulation box with
$l=66$\AA \  and cell size $6$\AA \ would scale over $167$
multiprocessors ($1336$ cores), or $6$ G200-class GPUs. 
 Figure \ref{fig:scalinglimit} shows the runtime of the inner and
outer non-bonded kernels on a water box as a function of block count per kernel
invocation. The minimum computation time for the fully-parallel case would be
$3.4$ ms/step on current hardware.  To further improve performance, optimisation of
the kernel or further subdivision of the computation would be required. 

\section{ Conclusions}

We have presented a molecular dynamics application, ACEMD, designed to reach the
microsecond timescale even on cost-effective workstation hardware using the computational 
power of GPUs. It supports the
CHARMM27 force field and Amber force field in CHARMM format and is
therefore suitable for use in modelling biomolecular systems.  The ability to
model these systems for tens of nanoseconds per day makes it feasible to
perform simulations of up to the microsecond scale over the course of a few
weeks on a suitable GPU-equipped machine.  Calculations lasting a few weeks are
perfectly reasonable tasks on workstation-class computers equipped with single
or multiple GPUs.  The current implementation limits the number of atoms to a
maximum of 250,000 atoms which could be extended.

\ACEMD has been extensively tested since August 2008 through its deployment on the several
thousand GPU-equipped PCs which participate in the volunteer distributed
computing project GPUGRID.net \cite{gpugridweb},
based on the Berkeley Open Infrastructure for Networked Computing (BOINC)
\cite{boinc} middleware.  At the time of writing, GPUGRID.net delivers over
$30$ Tflops of sustained performance \cite{boincstats}, and is thus one of the largest distributed
infrastructures for molecular simulations, producing thousands of
nanosecond long trajectories per day for high-throughput molecular
simulations, for instance for accurate virtual screening \cite{science}.

The current implementation of \ACEMD limits its parallel performance to just $3$ GPUs due to a simple task parallelization. 
We plan to extend the use of \ACEMD on more GPUs, but keeping the focus on scalability, so small numbers of GPUs (1-32).
Ideally the optimal system for \ACEMD  would rely on a single node attached to
a large number of GPUs via individual PCIe expansion slots in order to take advantage of
the large interconnect bandwidth. \ACEMD  would potentially scale very well on such machine
due to the fact that it is entirely executing on the GPU devices, obtaining CPU
loads within just 5\%.  For efficient scaling across a GPU-equipped cluster, we
anticipate that a re-factoring of the parallelisation scheme to use a spatial
decomposition method\cite{nt} would be necessary, moving away from the simple
task parallelisation used in this work.  Possible future developments also
include support of forthcoming programming languages for GPUs, for example
OpenCL \cite{opencl},
 a development library which is
intended to provide a hardware-agnostic, data-parallel programming model.
Whilst GPU devices are commonly present in desktop and
workstation computers for graphics purposes, as accelerator processors they
have yet to become routinely integrated components of the compute cluster
systems typically used for high-performance computing (HPC) systems.  GPU
workstations,  such as the one used in this work are readily available, while
GPU clusters are slowly appearing \footnote{\NVIDIA Tesla-equipped Tsubame
cluster, Tokyo Institute of Technology}. In order to scale efficiently,
low-latency, high-bandwidth communications between nodes is necessary. For
example, Bowers \textit{et al} \cite{Desmond}, describe the scaling of the
Desmond MD program over an Infiniband\cite{infiniband} network and demonstrate
improved scaling when using custom communications routines tailored to the
requirements of the algorithm and the capabilities of the network technology.

 Accelerated molecular dynamics on GPUs as provided by \ACEMD  should be of wide
interest to a large number of computational scientists as it provides performance 
comparable to that achievable on standard CPU supercomputers in a laboratory
environment. Even research groups that have routine access 
supercomputing time might find useful the ability to run simulations locally for longer
time windows and with added flexibility.

 
{\bf Acknowledgements.} This work was partially funded by the HPC-EUROPA
project (R113-CT-2003-506079). GG acknowledges support from the Marie Curie
Intra-European Fellowship (IEF).  GDF acknowledges support from the Ramon y
Cajal scheme and also from the EU Virtual Physiological Human Network of
Excellence.  We gratefully acknowledge the advice of Sumit Gupta (Nvidia),
Acellera Ltd (http://www.acellera.com) for use of their resources and \NVIDIA
Corporation (http://www.nvidia.com) for their hardware donations. We thank
Ignasi Buch for help in debugging the software.
\bibliography{../bibliography}
\bibliographystyle{achemso}

\end{document}